\documentclass[aps,prc,showpacs,amssymb,amsmath,amsfonts,twocolumn,floatfix,nofootinbib]{revtex4-1}
\usepackage{graphicx}
\usepackage{natbib}
\usepackage[usenames]{color}
\usepackage{epsfig}
\usepackage{amssymb,amsmath}
\usepackage{subfig}
\usepackage{lineno,hyperref}

\begin{document}

\title{Baryon transition form factors at the pole}

\author{L. Tiator$\,^1$, M. D\"{o}ring$\,^{2,3}$, R.~L.~Workman$\,^2$,
M. Had\v{z}imehmedovi\'{c}$\,^4$, H. Osmanovi\'{c}$\,^4$,
R. Omerovi\'{c}$\,^4$, J. Stahov$\,^4$, A. \v{S}varc$\, ^5$ }

\vspace{0.5 cm}

\affiliation{$^1$ Institut f\"{u}r Kernphysik, Universit\"{a}t Mainz,
                  D-55099 Mainz, Germany}
\affiliation{$^2$ The George Washington University, Washington, DC 20052, USA}
\affiliation{$^{3}$ Thomas Jefferson National Accelerator Facility, Newport News, VA 23606, USA}

\affiliation{$^4$ University of Tuzla, Faculty of Science, Univerzitetska 4,
                  75000 Tuzla, Bosnia and Herzegovina}
\affiliation{$^5$ Rudjer Bo\v{s}kovi\'{c} Institute, Bijeni\v{c}ka cesta 54,
                 P.O. Box 180, 10002 Zagreb, Croatia}
\date{\today}

\vspace{5cm}
\date{\today}

\begin{abstract}
Electromagnetic resonance properties are uniquely defined at the pole and do not depend on the
separation of the resonance from background or the decay channel. Photon-nucleon branching ratios
are nowadays often quoted at the pole, and we generalize the considerations to the case of virtual
photons. We derive and compare relations for nucleon to baryon transition form factors both for the
Breit-Wigner and the pole positions. Using the MAID2007 and SAID SM08 partial wave analyses of pion
electroproduction data, we compare the $G_M$, $G_E$, and $G_C$ form factors for the
$\Delta(1232)$ resonance excitation at the Breit-Wigner resonance and pole positions up to
$Q^2=5$~GeV$^2$. We also explore the $E/M$ and $S/M$ ratios as functions of $Q^2$. For pole and
residue extraction, we apply the Laurent + Pietarinen method.
\end{abstract}

\maketitle

\section{Introduction}
\label{sec:intro}

As baryon resonance properties, evaluated at the pole position, are now beginning to supersede and
replace quantities that historically have been determined using Breit-Wigner (BW) plus background
parameterizations, we extend a recent study~\cite{Workman:2013rca} of photo-decay couplings at the
pole to the regime of non-zero $Q^2$. The shift to pole-related quantities is reflected in the
Review of Particle Properties (RPP)~\cite{pdg}, with many pole values coming from the Bonn-Gatchina
multi-channel analyses~\cite{bnga}. Some plots of transition form factors at the BW position, as a
function of photon virtuality, are now also available in the RPP.

Calculations at the pole are, in principle, well-defined and less model-dependent than the BW
approach. However, the continuation of fit amplitudes to the pole is itself a possible source of
error. With the aim to minimize model dependence of the pole extraction procedure this uncertainty
has motivated numerous studies involving speed plots,  regularization methods, contour
integration~\cite{Doring:2009yv, ceci11, masj11, yang11, tiat10, suzu10}, and the most recent
Laurent series representations (L+P) based on separation of pole and regular part, and using the
conformal mapping variable to expand the regular part in power series \cite{svar12}.

While analysis in terms of BW amplitudes is more in line with the early fits, based on simple BW
plus background parameterizations, fits done within dynamical models are more naturally presented
in terms of poles~\cite{Kamano:2013iva, Ronchen:2015vfa, Ronchen:2014cna}. In the chiral effective
field theory calculation of Ref.~\cite{Gail}, for example, complex form factor results at the pole
could only be compared to real BW values coming from fits to data. See also
Refs.~\cite{Pascalutsa:2005vq, Bernard:2005fy} for related results. Also, in chiral unitary
resonance dynamics, there is no genuine resonance seed that would allow for the definition of a
meaningful, purely real helicity coupling~\cite{Jido:2007sm, Doring:2010rd, Doring:2007rz}. As a
result, ad-hoc prescriptions were used to translate these results to quoted real BW helicity
amplitudes. A proper comparison, however, requires the associated quantities extracted at
the pole of amplitudes obtained from fits to experimental data.

To achieve this goal, we have extracted pole parameters with the L+P method from two
energy-dependent (ED) partial wave analyses, MAID and SAID, which are fitted to the world data base
of pion electroproduction. The differences observed by this comparison will give an insight of the
uncertainty of the pole form factors due to the differences in the MAID and SAID techniques and
consequently also to the experimental database.

For the $\Delta (1232)$ state, considerable attention~\cite{Aznauryan:2016wwm, Ramalho:2015qna,
aznauryan1, segovia, Segovia:2013uga,  Ronniger:2012xp, Sanchis-Alepuz:2013iia, Ramalho:2012ng,
Santopinto:2012nq, Aznauryan:2011qj, Santopinto:2010zz, ebac, ungaro, Fiolhais:1996bp} has been
addressed to the $Q^2$ evolution of amplitudes, as well as differences in the $Q^2$ dependence of
bare couplings, within models, and meson-cloud contributions. For a review, see, e.g.,
Ref.~\cite{Pascalutsa:2006up}. Transition form factors are now also determined in lattice QCD
simulations~\cite{Alexandrou:2013ata, Alexandrou:2010uk}. Here, the quark masses are so large that
the $\Delta(1232)$ appears as a bound state, but it has been realized in
Ref.~\cite{Agadjanov:2014kha} that in future simulations, close to the physical point, the finite
resonance width will complicate the extraction. Therefore, in Ref.~\cite{Agadjanov:2014kha} a
method has been proposed to determine transition form factors at the pole. This stresses again the
relevance of providing pole values for existing phenomenological analyses, which is the aim of this
study.

The main focus of this paper is the $\Delta (1232)$ virtual-photon decay amplitudes
and related transition form factors.
As the amplitudes themselves become infinite at the pole, we are interested in residues.
The connection between multipole residues and the photo-decay amplitudes has been clarified in
a previous paper~\cite{Workman:2013rca} and in the Note on N and Delta Resonance mini-reviews of the
2012 and 2014 PDG listings~\cite{pdg}.

Here we will compare BW and pole extractions, using the MAID2007 and SAID SM08 partial wave
analyses of pion electroproduction data, utilizing the recent Laurent + Pietarinen (L+P) pole
extraction method \cite{L+P2013,L+P2014,L+P2014a,L+P2015,L+P2016} which has proven to be a precise
and very reliable tool for the determination of pole positions and residues.

The $\Delta (1232)$ E/M and S/M ratios have been studied for many decades. Interest in the E/M
ratio, for real-photon interactions, was largely motivated by the fact that, in a simple non-relativistic
quark
model, this ratio would be zero~\cite{bm} and, thus, deviations from zero would require more
complicated interactions. The measured value of this ratio was small~\cite{pdg}, $-2.5\pm 0.5 \%$,
but its precise value varied
as photoproduction cross sections and beam-asymmetry $\Sigma$ measurements became more precise. The
prediction for this ratio, at very large $Q^2$ from pQCD~\cite{carlson}, has been more difficult to
confirm. The ratio is predicted to become unity, whereas, at the real-photon point, it
is small and negative. The variation of this ratio with $Q^2$ has also changed significantly as
electroproduction data have improved.

In Section II, we first give definitions of the standard BW quantities and then define the
associated pole-valued results we will be considering. In Section III, we then give a brief overview
of the BW and pole behavior of the $\Delta (1232)$ amplitudes, which are constrained by Watson's theorem.
The L+P fit is described in Section IV and compared to results from a fit, generalized to non-zero $Q^2$,
described in Ref.~\cite{Workman:2013rca}. Finally, in Section IV, we summarize our findings
and prospects for future work.

\section{Breit-Wigner versus pole quantities}
\label{sec:said90}

The total cross section of pion electroproduction can be written as a semi-inclusive electron
scattering cross section
\begin{equation}
\frac{d\sigma}{d\Omega_e dE_f}=\Gamma_V(\sigma_T+\varepsilon\sigma_L)
\end{equation}
with the virtual-photon flux factor
\begin{equation}
\Gamma_V=\frac{\alpha}{2\pi^2}\frac{E_f}{E_i}\frac{\kappa_\ell}{Q^2}\frac{1}{1-\varepsilon}\,,
\end{equation}
where $E_i$ and $E_f$ are the initial and final electron energies in the lab frame, the
virtual-photon polarization $\varepsilon$ and the total transverse and longitudinal virtual-photon
cross sections
\begin{eqnarray}
\sigma_{T}&=&\frac{1}{2} (\sigma_{T}^{1/2}+\sigma_{T}^{3/2})\,,\\
\sigma_{T}^{h}&=&4\pi\frac{q}{\kappa_c}\sum_{\alpha(\ell,J,I)}
(2J+1)\,|\mathcal{A}_\alpha^{h}|^2\;C^2\,,\\
\sigma_{L}&=&4\pi\frac{q}{\kappa_c}\frac{Q^2}{k^2}\sum_{\alpha(\ell,J,I)}
(2J+1)\,|\mathcal{S}_\alpha^{1/2}|^2\;C^2\,, \label{sigtot}
\end{eqnarray}
with $q$ and $k$ being the center-of-mass pion and photon momenta and $\kappa_l=(W^2-m_N^2)/2m_N$
and $\kappa_c=(W^2-m_N^2)/2W$ the so-called equivalent real photon energies in the lab and c.m.
frames. The factor $C$ is $\sqrt{2/3}$ for isospin $3/2$ and $-\sqrt{3}$ for isospin $1/2$.  The
helicity multipoles are given in terms of electric, magnetic and longitudinal (time-like)
multipoles
\begin{eqnarray}
\mathcal{A}_{\ell +}^{1/2} & = & -{1\over 2} \left[ (\ell +2)
{E}_{\ell +}
+ \ell {M}_{\ell +} \right] ,\label{helimult1}
\\
\mathcal{A}_{\ell +}^{3/2} & = & {1\over 2} \sqrt{\ell ( \ell + 2)} \left[ {E}_{\ell +} -{M}_{\ell
+} \right] , \label{helimult2}
\\
\mathcal{S}_{\ell +}^{1/2} & = & -{\ell+1\over \sqrt{2}} {S}_{\ell +}\,, \label{helimult2}
\\
\mathcal{A}_{(\ell +1)-}^{1/2} & = & -{1\over 2} \left[\ell {E}_{(\ell +1)-} - (\ell +2 )
{M}_{(\ell +1)-} \right] , \label{helimult3}
\\
\mathcal{A}_{(\ell +1)-}^{3/2} & = & -{1\over 2} \sqrt{\ell ( \ell +2)} \left[ {E}_{(\ell +1)-} +
{M}_{(\ell +1) -} \right] ,\\
\mathcal{S}_{(\ell +1)-}^{1/2} & = & -{\ell+1\over \sqrt{2}} {S}_{(\ell +1)-}\,, \label{helimult4}
\end{eqnarray}
with $J=\ell+1/2$ for '$+$' multipoles and $J=(\ell+1)-1/2$ for
'$-$' multipoles, all having the same total spin $J$.

In analogy to photoproduction~\cite{Workman:2013rca}, we define the virtual-photon decay amplitudes
\begin{eqnarray}
A_h^{BW}&=&C\,\sqrt{\frac{q_r}{\kappa_r}\frac{\pi(2J+1)M_r\Gamma_r^2}{m_N\Gamma_{\pi,r}}}\;
\tilde{\mathcal{A}}_\alpha^h\,,\\
S_{1/2}^{BW}&=&C\,\sqrt{\frac{q_r}{\kappa_r}\frac{\pi(2J+1)M_r\Gamma_r^2}{m_N\Gamma_{\pi,r}}}\;
\tilde{\mathcal{S}}_\alpha^{1/2}\,,
\end{eqnarray}
where
$\tilde{\mathcal{A}}_\alpha^{1/2}$, $\tilde{\mathcal{A}}_\alpha^{3/2}$, and $\tilde{\mathcal{S}}_\alpha^{1/2}$
are the imaginary parts of the resonance amplitudes at the BW position $W_r=M_r$.

Similarly, we define the virtual-photon amplitudes at the pole position
\begin{eqnarray}
A_h^{pole}&=&C\,\sqrt{\frac{q_p}{\kappa_p}\frac{2\pi(2J+1)W_p}{m_N Res_{\pi N}}}\;
\mbox{Res}\,\mathcal{A}_\alpha^h\, ,\\
S_{1/2}^{pole}&=&C\,\sqrt{\frac{q_p}{\kappa_p}\frac{2\pi(2J+1)W_p}{m_N Res_{\pi N}}}\;
\mbox{Res}\,\mathcal{S}_\alpha^{1/2}\, .
\end{eqnarray}
where the subscript $p$ denotes quantities evaluated at the pole
position. 

The photon momenta, $\kappa_r$ and $\kappa_p$, are photon equivalent energies and can be written as
virtual-photon momenta at $Q^2=0$. The amplitudes, $\mathcal{A}_\alpha^h$ and
$\mathcal{S}_\alpha^{1/2}$, as well as the residues, $A_h^{pole}$ and $S_{1/2}^{pole}$, are
functions of the photon virtuality $Q^2$.

Through linear combinations, the helicity form factors can also be related to electric, magnetic and
charge form factors. These so-called Sachs form factors, $G_E^*$, $G_M^*$, and $G_C^*$, are usually given in
two different conventions by Ash~\cite{Ash67} and by Jones and Scadron~\cite{Jon73}. Both are related
by a square-root factor, $G_{JS}(Q^2)=G_{Ash}(Q^2)\times\sqrt{1+Q^2/(m_N+M_r)^2}$.

Here we will concentrate on the $\Delta(1232)$ transition form factors and give the corresponding
expressions. For transitions with different spin and parity, similar relations can be found, see
\cite{Devenish:1975jd,Tiator:2011pw}.

For the $\gamma N \Delta$ transition, Jones and Scadron~\cite{Jon73} give the following relations
between the total cross sections and the Sachs form factors:
\begin{eqnarray}
\sigma_T+\varepsilon\sigma_L &=& \frac{2\pi\alpha}{\Gamma_r m_N^2}
\frac{k_r(Q^2)^2}{\kappa_r(1+Q^2/(m_N+M_\Delta)^2}\\
&&({G_M^*}^2(Q^2) + 3 {G_E^*}^2(Q^2) + \varepsilon\frac{Q^2}{4M_\Delta^2} {G_C^*}^2(Q^2))\,. \nonumber
\end{eqnarray}

In the convention of Ash, the Sachs form factors take the form
\begin{align}
G_M^*(Q^2) &=  -c_\Delta (A_{1/2}+\sqrt{3} A_{3/2})\,,\\
G_E^*(Q^2) &=  \,\,\, c_\Delta (A_{1/2}-\frac{1}{\sqrt{3}} A_{3/2})\,,\\
G_C^*(Q^2) &= \sqrt{2} c_\Delta \frac{2M_\Delta}{k_\Delta} S_{1/2}\,,
\end{align}
with $W_r=M_\Delta$ and $c_\Delta = [ ( m_N^3 \kappa_\Delta )/ (4\pi\alpha M_\Delta k_\Delta^2
)]^{1/2}$, where  $k_\Delta=k_\Delta (Q^2)=k(M_{\Delta},Q^2)$, and
$\kappa_\Delta=\kappa_c(M_\Delta)=k(M_\Delta,0)$ are the virtual-photon momentum and the photon
equivalent energy at resonance. Because the $\Delta(1232)$ is very close to an ideal resonance, the
real parts of the amplitudes vanish at $W=M_{\Delta}$, and the form factors can be directly
expressed in terms of the imaginary parts of the corresponding multipoles at the (Breit
Wigner) resonance position,
\begin{align}
G^{\ast}_M(Q^2) &= b_\Delta(Q^2)\, {\rm {Im}}\{M_{1+}^{(3/2)}(M_\Delta,Q^2)\}
\label{eq:GstarMtoMult}\\
G^{\ast}_E(Q^2) &= -b_\Delta(Q^2)\, {\rm {Im}}\{E_{1+}^{(3/2)}(M_\Delta,Q^2)\}
\label{eq:GstarEtoMult}\\
G^{\ast}_C(Q^2) &= -b_\Delta(Q^2) \frac{2M_{\Delta}}{k_{\Delta}(Q^2)}\,
{\rm {Im}}\{S_{1+}^{(3/2)}(M_\Delta, Q^2)\}\,,\label{eq:GstarCtoMult}
\end{align}
where $b_\Delta(Q^2) = [ ( 8 m_N^2 q_\Delta \Gamma_\Delta )/ (3 \alpha
k_\Delta^2(Q^2) )]^{1/2}$
with $\Gamma_{\Delta}=115$ MeV and $q_{\Delta}=q(M_{\Delta})$ being the pion momentum at resonance.

Similarly, we can define the Sachs form factors, using Ash's conventions, at the pole position
\begin{align}
G^{pole}_M(Q^2) &= b_p(Q^2)\, {\mbox{Res}}\,M_{1+}^{(3/2)}(W_p,Q^2)
\,,\label{eq:GpoleMtoMult}\\
G^{pole}_E(Q^2) &= -b_p(Q^2)\, {\mbox{Res}}\,E_{1+}^{(3/2)}(W_p,Q^2)
\,,\label{eq:GpoleEtoMult}\\
G^{pole}_C(Q^2) &= -b_p(Q^2) \frac{2W_p}{k_p(Q^2)}\, {\mbox{Res}}\,S_{1+}^{(3/2)}(W_p,
Q^2)\,,\label{eq:GpoleCtoMult}
\end{align}
where $b_p(Q^2) = [ ( 16\, m_N^2\, q_p )/(3\,\alpha\, k_p^2(Q^2)\,Res_{\pi N} )]^{1/2}$.

In the literature, the following ratios of multipoles have been defined:
\begin{eqnarray}
R_{EM}(Q^2) &=& \frac{E_{1+}^{3/2}(Q^2)}{M_{1+}^{3/2}(Q^2)} =
-\frac{G_E^{\ast}(Q^2)}{G_M^{\ast}(Q^2)}\,,\label{eq:REM}\\
R_{SM}(Q^2) &=& \frac{S_{1+}^{3/2}(Q^2)}{M_{1+}^{3/2}(Q^2)} =
-\frac{k_\Delta(Q^2)}{2M_{\Delta}}\frac{G_C^{\ast}(Q^2)}{G_M^{\ast}(Q^2)}\,. \label{eq:RSM}
\end{eqnarray}
The ratios at the pole position are given accordingly
\begin{eqnarray}
R_{EM}^{pole}(Q^2) &=& \frac{\mbox{Res}\,E_{1+}^{3/2}(Q^2)}{\mbox{Res}\,M_{1+}^{3/2}(Q^2)} =
-\frac{G_E^{pole}(Q^2)}{G_M^{pole}(Q^2)}\,,\label{eq:REM}\\
R_{SM}^{pole}(Q^2) &=& \frac{\mbox{Res}\,S_{1+}^{3/2}(Q^2)}{\mbox{Res}\,M_{1+}^{3/2}(Q^2)} =
-\frac{k_p(Q^2)}{2W_p}\frac{G_C^{pole}(Q^2)}{G_M^{pole}(Q^2)}\,. \label{eq:RSM} \nonumber \\
\end{eqnarray}

\section{Amplitudes at the Breit-Wigner position and at the pole position}
\label{sec:amplitudes}

In general, a pion photo- or electroproduction amplitude $T_{\gamma,\pi}$, or any multipole, can be
written as a sum of resonance and background contributions
\begin{equation}
T_{\gamma,\pi}(W,Q^2) = T_{\gamma,\pi}^{res}(W,Q^2)+ T_{\gamma,\pi}^{bg}(W,Q^2)\,.
\end{equation}
In order to obtain the BW amplitudes at the resonance position, the resonance part has to be
modeled in BW form with energy-dependent partial widths for all possible decay channels and with
energy-dependent phases in order to obey unitarity, see e.g. Ref.~\cite{maid2007}. In general, this
resonance-background separation is only possible in a model dependent way~\cite{Doring:2009bi}.
Consequently, this also leads to some model dependence in the mass $M_r$, width $\Gamma_r$ and for
the amplitudes $\mathcal{A}_\alpha^h$ and $\mathcal{S}_\alpha^{1/2}$. The only exception in the
baryonic spectrum is the $\Delta(1232)$, which is purely elastic and therefore has a well-defined
K-matrix pole, $M_\Delta=1232$~MeV, where the scattering phase $\delta_{33}(M_\Delta)=\pi/2$. For
the $\Delta(1232)$ this coincides with the Breit-Wigner resonance position. Due to Watson's
theorem, also for pion photo- and electroproduction, the phase is exactly 90 degrees at resonance
and the resonance-background separation is unique, as the background amplitude
$T_{\gamma,\pi}^{bg}(M_\Delta,Q^2)=0$.

This situation is very different at the pole position. Since only the resonance part of the total
amplitude contains a pole, a model-dependent resonance-background separation is unnecessary.
Therefore, the pole positions and also the residues are model independent. However, they can suffer
from uncertainties arising from the analytical continuation of the amplitudes, determined from data
on the real energy axis, into
the lower part of the complex energy plane.

\section{L+P expansion}
\label{sec:L+P}

Finding the pole positions and the residues of baryon resonances can be a difficult task. Some
first attempts, applied to $\pi N$ scattering amplitudes, were carried out
by H\"ohler~\cite{Hoehler93} and Cutkosky~\cite{Cutkosky79}.
The optimal method would be an analytic continuation into the
complex plane, within a dynamical and analytical model, carefully considering all branch cuts from
open channels, that generally produce many poles on different Riemann sheets, where only
the pole closest to the physical axis is relevant. In many cases, however, this is not
possible in practice, e.g. when partial wave amplitudes can only be evaluated on the physical axis.
For these cases,
H\"ohler proposed the speed-plot technique~\cite{Hoehler93}, which was later extended by the regularization
method~\cite{Ceci2008}. In the present study,
we apply the Laurent-plus-Pietarinen method (L+P)  based on separation of pole and regular part, and using the conformal mapping variable to expand the regular part in power series; the method which has proven to be most
reliable and has been applied to different
processes~\cite{L+P2013,L+P2014,L+P2015,L+P2014a,L+P2016}. A major advantage of the L+P
method is the fact that it is a global method, describing the amplitudes over a wide energy range,
treating threshold effects in terms of physical and effective branch points. Most other methods use
only partial wave information in a local region around the relevant resonance position.

In this study, we have adopted the multi-channel Laurent-plus-Pietarinen method (MC L+P), developed
in Ref.~\cite{L+P2016}, to the single-channel case where the $E$, $M$, and $S$ multipoles must be
treated simultaneously as they share the same pole with associated $\pi N$ resonance coupling. One
could therefore describe the method as a coupled-multipole Laurent-plus-Pietarinen method (CM L+P).

\begin{widetext}
\begin{eqnarray}
\label{eq:Laurent-Pietarinen}
T_{\gamma,\pi}^a(W) &=& \sum _{j=1}^{{N}_{pole}} \frac{x^{a}_{j} + \imath \, \, y^{a}_{j}  }{W_j-W} + \nonumber \\
       & + & \sum _{k=0}^{K^a}  c^a_k \, X^a (W)^k  +  \sum _{l=0}^{L^a} d_l^a \, Y^a (W)^l +  \sum
_{m=0}^{M^a} e_m^a \, Z^a (W)^m \nonumber \\
X^a (W )&=& \frac{\alpha^a-\sqrt{x_P^a-W}}{\alpha^a+\sqrt{x_P^a - W }}; \, \, \, \, \,   Y^a(W ) =  \frac{\beta^a-\sqrt{x_Q^a-W }}{\beta^a+\sqrt{x_Q^a-W }};  \, \, \, \, \,
Z^a(W ) =  \frac{\gamma^a-\sqrt{x_R^a-W}}{\gamma^a+\sqrt{x_R^a-W }}
 \nonumber \\
 D_{dp} & = & \sum _{a}^{all}D_{dp}^a \nonumber \\
 D_{dp}^a & = & \frac{1}{2 \, N_{W}^a} \, \, \sum_{i=1}^{N_{W}^a} \left\{ \left[ \frac{{\rm Re} \,T _{\gamma,\pi}^{a}(W^{(i)})-{\rm Re} \, T_{\gamma,\pi}^{a,exp}(W^{(i)})}{ Err_{i,a}^{\rm Re}}  \right]^2 + \right. \nonumber \\
 &+&    \left. \left[ \frac{{\rm Im} \, T_{\gamma,\pi}^{a}(W^{(i)})-{\rm Im} \, T_{\gamma,\pi}^{a,exp}(W^{(i)})}{ Err_{i,a}^{\rm Im}} \right]^2 \right\}  + {\cal P}^a  \nonumber \\
 Err_{i,a}^{{\rm Re}}&=&0.05\cdot\dfrac{\sum_{k=1}^{N_W^a}|{\rm Re\;}T_{\gamma,\pi}^a(W^{(k)})|}{N_W^a}+0.05\cdot|{\rm Re\;}T_{\gamma,\pi}^a(W^{(i)})|\nonumber\\
Err_{i,a}^{{\rm Im}}&=&0.05\cdot\dfrac{\sum_{k=1}^{N_W^a}|{\rm Im\;}T_{\gamma,\pi}^a(W^{(k)})|}{N_W^a}+0.05\cdot|{\rm Im\;}T_{\gamma,\pi}^a(W^{(i)})| \nonumber \\
{\cal P}^{a} &=& \lambda_c^a \, \sum _{k=1}^{K^a} (c^a_k)^2 \, k^3 +  \lambda_d^a \, \sum _{l=1}^{L^a} (d^a_l)^2 \, l^3 +
 \lambda_e^a \, \sum _{m=1}^{M^a} (e^a_m)^2 \, m^3 \nonumber \\
 && a \, \, .....   \, \,    {\rm  multipole \, \,  index}{ \, \, (E_{l_\pm}, M_{l_\pm}, S_{l_\pm})} \nonumber
\nonumber \\
 && N_{pole}\; .....\;      {\rm number\; of\; poles} \nonumber \\
 &&  W_j,W \in \mathbb{C}  \nonumber \\
&& x_i^a, \, y_i^a, \, c_k^a, \, d_l^a, \, e_m^a, \, \alpha^a, \, \beta^a, \, \gamma^a ... \in  \mathbb{R}  \, \,
 \nonumber \\
 &&  K^a, \, L^a, \, M^a \, ... \, \in  \mathbb{N} \, \, \, {\rm number \, \, of \, \, Pietarinen \, \, coefficients \, \, in \, \, multipole \, \, \mathit{a} }.
 \nonumber \\
 && N_{W}^a \; .....  \; {\rm number\; of\; energies\; in\; multipole \; }a \nonumber \\
 && {\cal P}^a \, \,  .....   \, \, {\rm Pietarinen  \, \, penalty \, \, function} \nonumber \\
 && \lambda_c^a, \, \lambda_d^a, \,\lambda_e^a \, \, \,  .....   \, \, {\rm Pietarinen  \, \,  weighting \, \, factors} \nonumber \\
&& Err_{i,a}^{\rm Re, \, Im} ..... {\rm \, \, minimization \, \, error\, \, of \, \, real \,\, and \, \, imaginary \, \, part \, \, respectively,} \nonumber \\
 && x_P^a, \, x_Q^a, \, x_R^a  \in \mathbb{R} \, \, \, \,( {\rm  or} \, \, \in \mathbb{C}).   \nonumber \\
\end{eqnarray}
\end{widetext}

Here $x_j^a + \imath \, \, y_j^a$ are the channel (multipole) residua which are left free for all
three multipoles $E$, $M$, and $S$ and \mbox{$W_j = M_j - \imath \, \, \frac{\Gamma_j}{2}$} are the
pole positions of resonances ``j'', which are kept fixed to the values obtained from L+P fits of
the single M-multipole obtained in the real photon case \cite{L+P2014a}). In addition, as we expect
a similar analytic structure for all three multipoles, we have fixed the three branch-points to
have the same value: $x_P^E=x_P^M=x_P^S$, $x_Q^E=x_Q^M=x_Q^S$ and $x_R^E=x_R^M=x_R^S$.

The Pietarinen expansions formalize the simplest analytic form of functions having branch-points at the
Pietarinen-expansion parameters, and in this paper we use three Pietarinen expansions with expansion
parameters, $x_P^a$, $x_Q^a$, and $x_R^a$, to describe analytic structure of the non-resonant background.

The first coefficient, $x_P^a$, is restricted to the unphysical region and effectively represents
contributions from all singular parts below $\pi N$ threshold (all left hand cuts including a
circular cut). The second parameter, $x_Q^a$, is fixed to the pion threshold at $W=1.077\;{\rm
GeV}$. The third branch-point, $x_R^a$, for MAID multipoles is left free and effectively accounts
for all inelastic-channel openings in the physical domain. Its values are above $\pi\pi N$
threshold. For SAID multipoles, $x_R^a$  is fixed at the complex branch-point $\pi
\Delta=(1.37-\imath\,0.04)\,{\rm GeV}$, and it effectively parameterizes all inelastic-channel
openings in the physical domain and a resonance in the three-body intermediate state.

In the fitting procedure we have used two poles for P$_{33}$ MAID amplitudes, and we used only one
pole and a complex branch-point for P$_{33}$ SAID amplitudes. However, as a complex branch-point
describes a pole hidden in a two-body channel of a three-body intermediate state, SAID is described
by two poles as well.

The L+P fit was compared to a method used to extract photo-decay amplitudes
at the pole in Ref.~\cite{Workman:2013rca}.  Residues were extracted from the SAID electroproduction
multipoles
for $Q^2$ from 0.1 to 5.0 GeV$^2$. Application to the SAID multipoles had the benefit of a known pole and
cut structure and a narrower range of fit energies was required. The values obtained in this way, and those
found using the L+P method, were not significantly different. This served as an independent test of the
L+P method applied to the electroproduction reaction. The L+P method was subsequently
used exclusively to obtain results from both the SAID and MAID multipoles.

\begin{figure*}[h]
\begin{center}
\includegraphics[width=16.0cm]{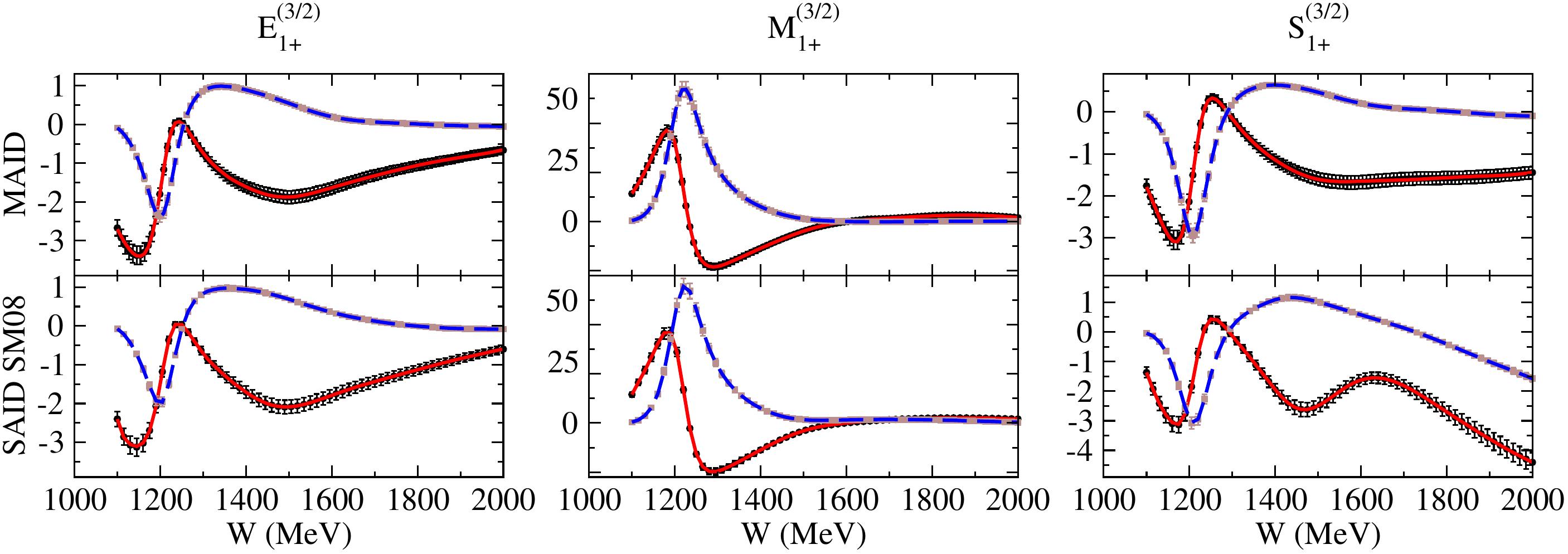} \\
\vspace*{1.cm}
\includegraphics[width=16.0cm]{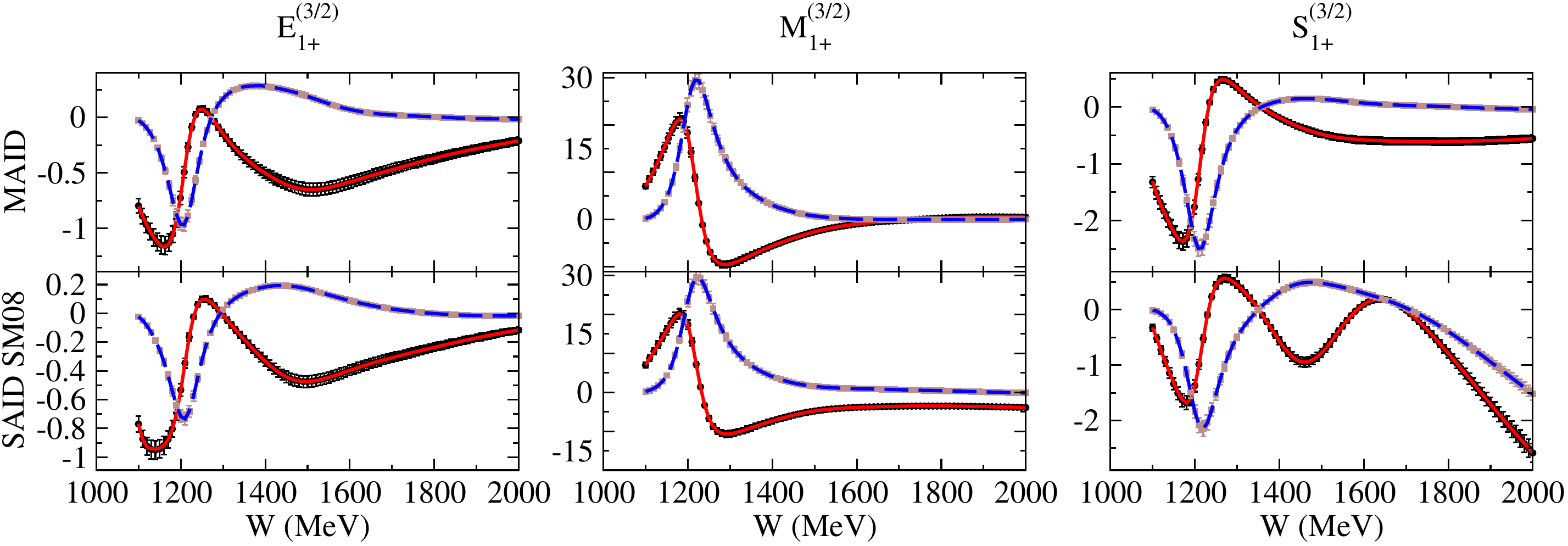} \\
\vspace*{1.cm}
\includegraphics[width=16.0cm]{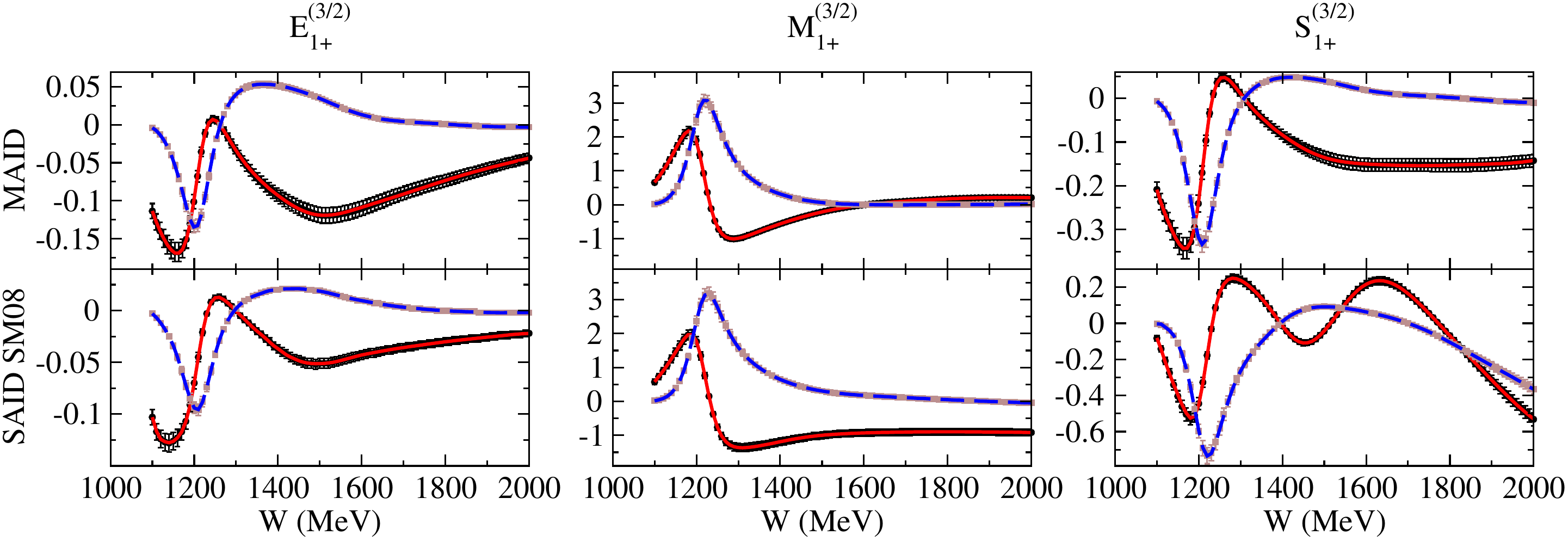}
\vspace{3mm} \caption{\label{fig:1} Figures showing the quality of the fit. We show all three
multipoles at three intermediate momentum transfers $Q^2=0$, 1, and 5 GeV$^2$ for MAID2007 and SAID
SM08 models. Black circles and brown squares are real and imaginary part of multipoles
respectively. Full red line is real part and blue dashed line is imaginary part of the L+P fit to
the given model.}
\end{center}
\end{figure*}

\section{Results and Conclusions}
\label{sec:results}

In the L+P analysis, the pion electroproduction multipoles $E_{1+}^{(3/2)}(W,Q^2)$,
$M_{1+}^{(3/2)}(W,Q^2)$, and $S_{1+}^{(3/2)}(W,Q^2)$, from MAID2007 and SAID SM08, were fitted from
threshold up to 2 GeV in the center-of-mass energy. These multipoles, which are
accessible via the MAID and SAID web pages~\cite{web}, are displayed in Fig. \ref{fig:1}. \\
For $Q^2$ values near the real-photon point, we fitted amplitudes from $Q^2=0$ to 0.5 GeV$^2$ in
increments of 0.1 GeV$^2$. We then examined $Q^2$ values in increments of 1 GeV$^2$ up to 5
GeV$^2$, finding this region to have a less rapid variation. At each value of $Q^2$, amplitudes
were analyzed in steps of 10 MeV.

Representative fit results covering the $\Delta (1232)$ energy range in Fig. \ref{fig:1} illustrate
the very good fit quality and also display the rapid fall off of these amplitudes with $Q^2$.
Numerical results from the $Q^2=0$ analyses are compiled in Table I.

\begin{figure*}[h]
\begin{center}
\includegraphics[width=7.5cm]{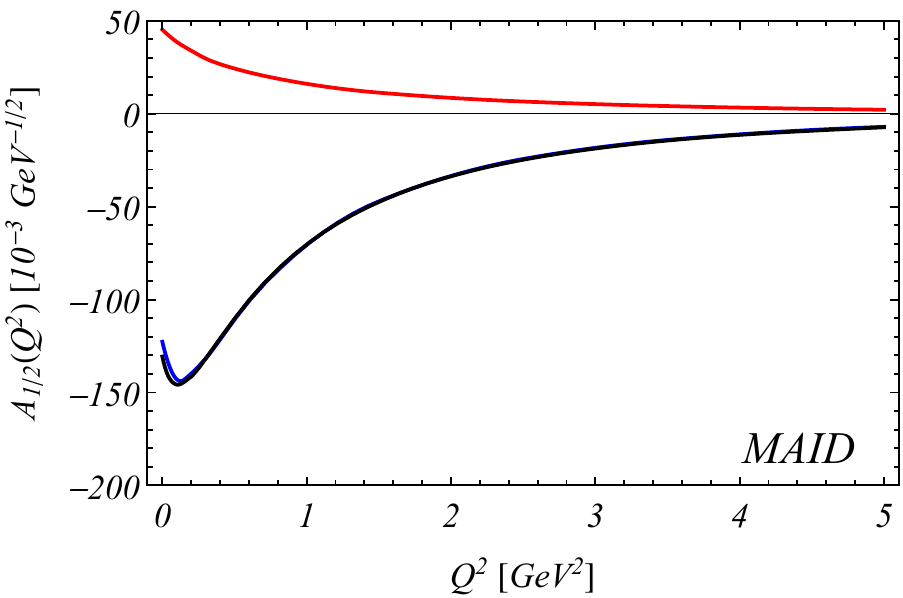}
\hspace*{0.5cm}
\includegraphics[width=7.5cm]{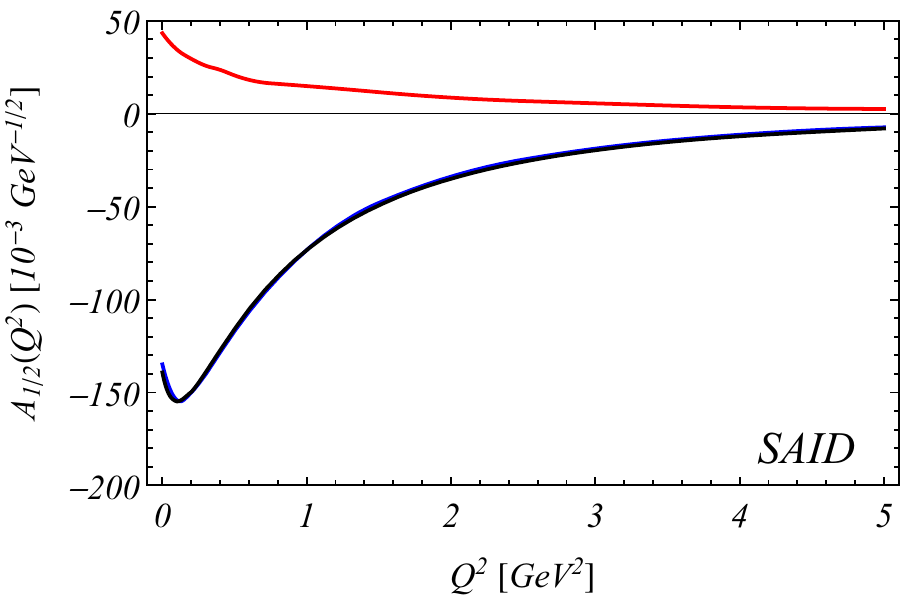}

\vspace*{0.5cm}
\includegraphics[width=7.5cm]{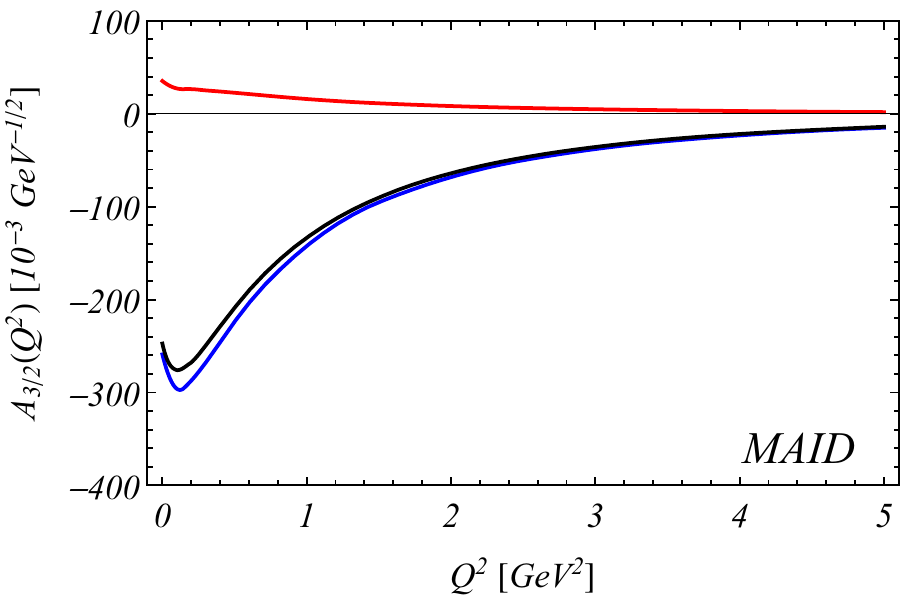}
\hspace*{0.5cm}
\includegraphics[width=7.5cm]{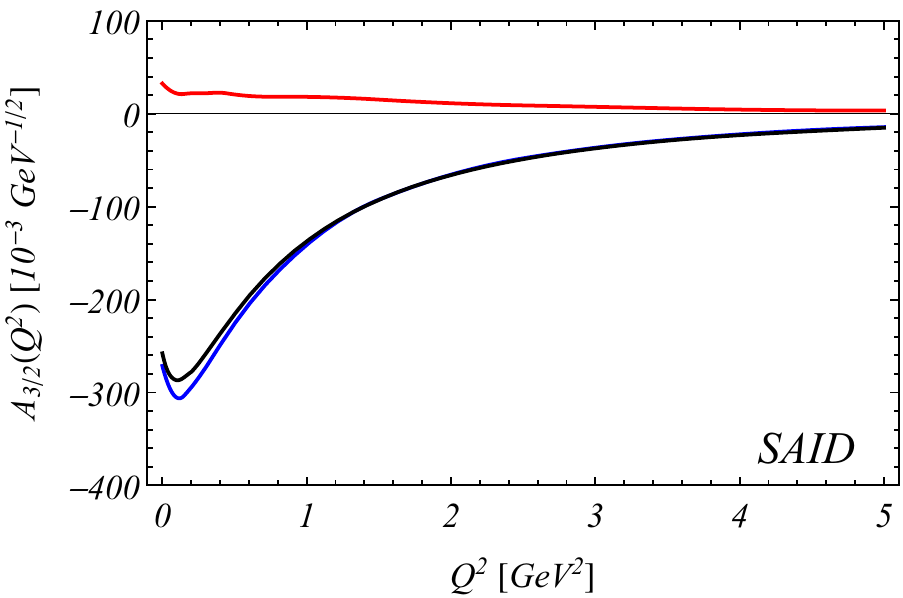}

\vspace*{0.5cm}
\includegraphics[width=7.5cm]{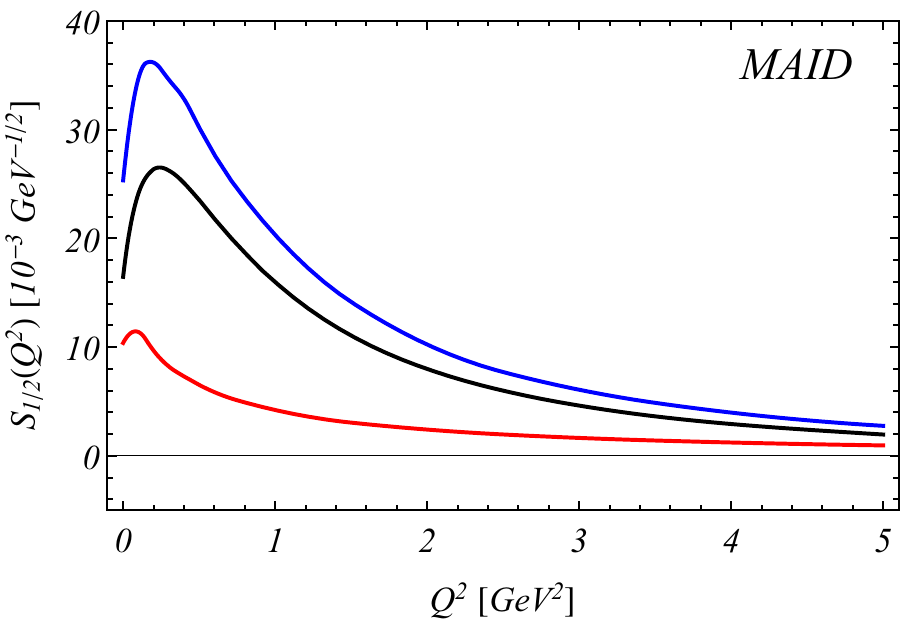}
\hspace*{0.5cm}
\includegraphics[width=7.5cm]{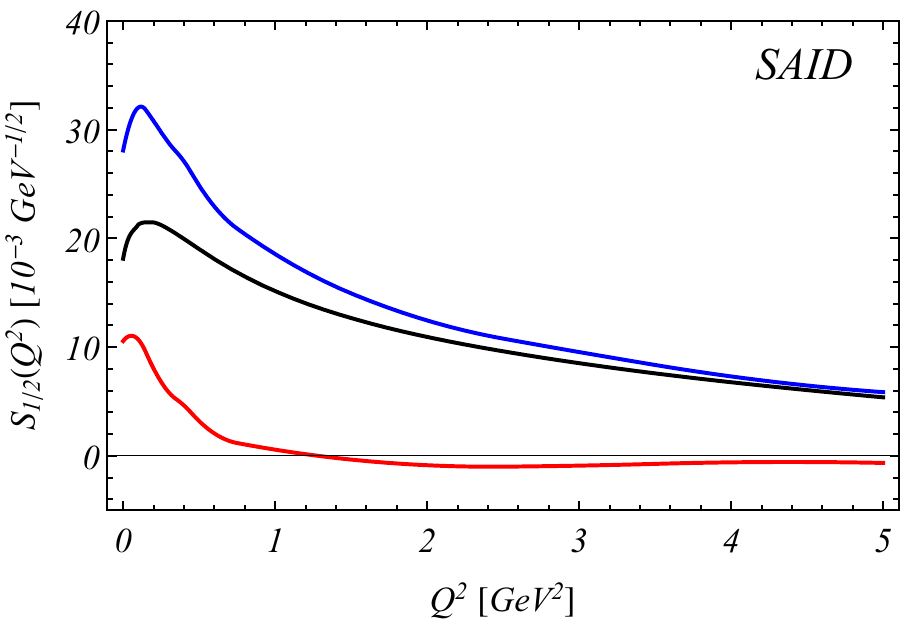}
\end{center}
\vspace{3mm} \caption{\label{fig:2} Helicity transition form factors $A_{1/2}$, $A_{3/2}$ and
$S_{1/2}$ compared at the BW and pole position. The black curves show the real BW results and the
blue and red lines show real and imaginary parts of the complex pole form factors. The left column
shows the results with the Mainz-MAID analysis and the right column with the GWU-SAID analysis.}
\end{figure*}

In Fig. \ref{fig:2}, we plot the associated helicity transition form factors, $A_{1/2}$, $A_{3/2}$,
and $S_{1/2}$ as functions of $Q^2$. The $A_{1/2}$ and $A_{3/2}$ amplitudes, being dominated by
the well-determined magnetic multipole, are very similar for the MAID and SAID analyses. The
$S_{1/2}$ variation in $Q^2$ is qualitatively similar but differs in detail. It is interesting to
note that, for the $A_{1/2}$ and $A_{3/2}$ amplitudes, the BW values and real parts of the pole quantities
are nearly identical, particularly with increasing $Q^2$.

\begin{figure*}[h]
\begin{center}
\includegraphics[width=7.5cm]{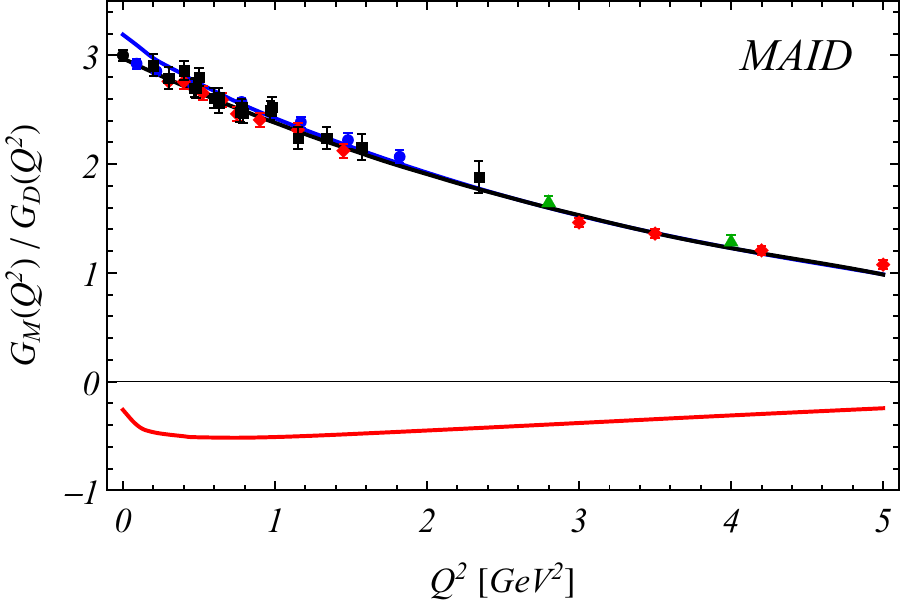}
\hspace*{0.5cm}
\includegraphics[width=7.5cm]{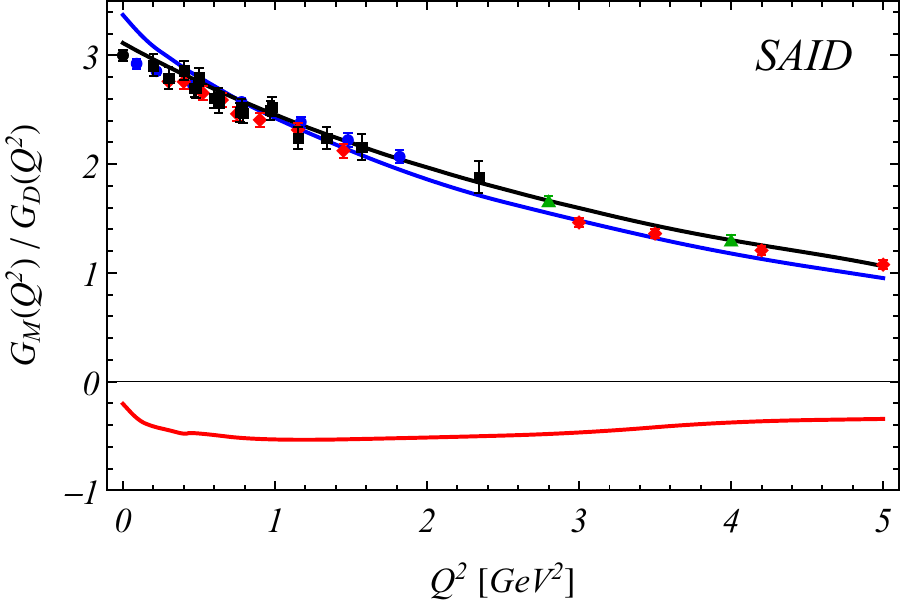}

\vspace*{0.5cm}
\includegraphics[width=7.5cm]{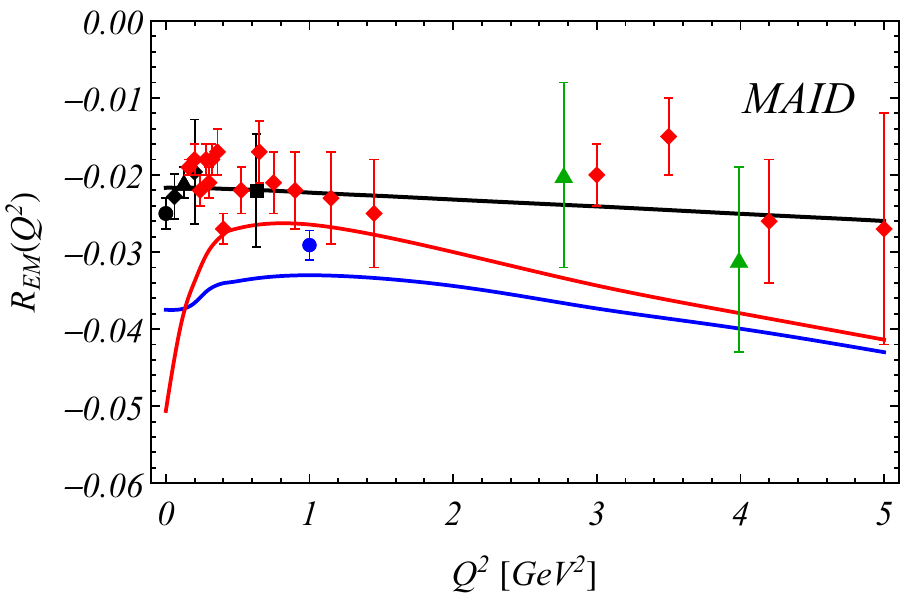}
\hspace*{0.5cm}
\includegraphics[width=7.5cm]{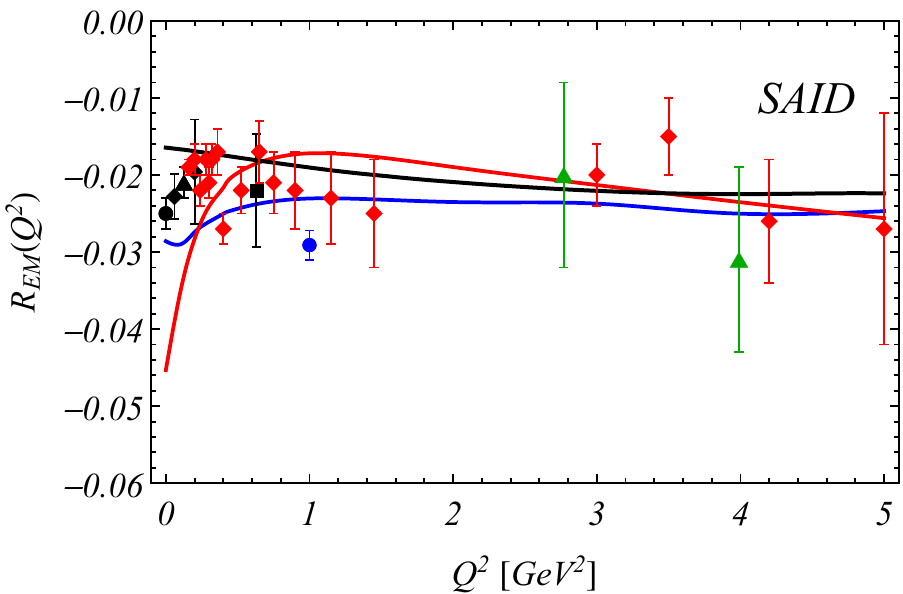}

\vspace*{0.5cm}
\includegraphics[width=7.5cm]{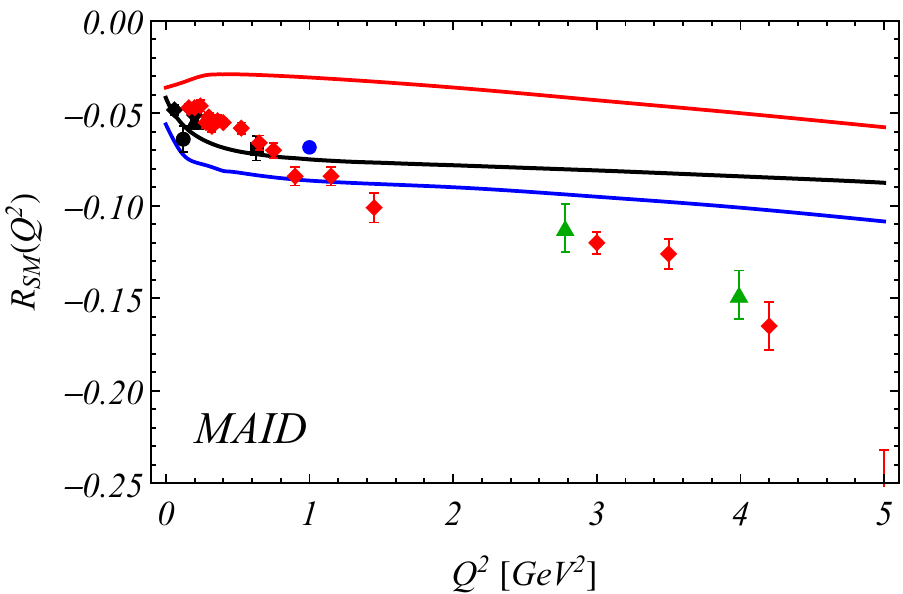}
\hspace*{0.5cm}
\includegraphics[width=7.5cm]{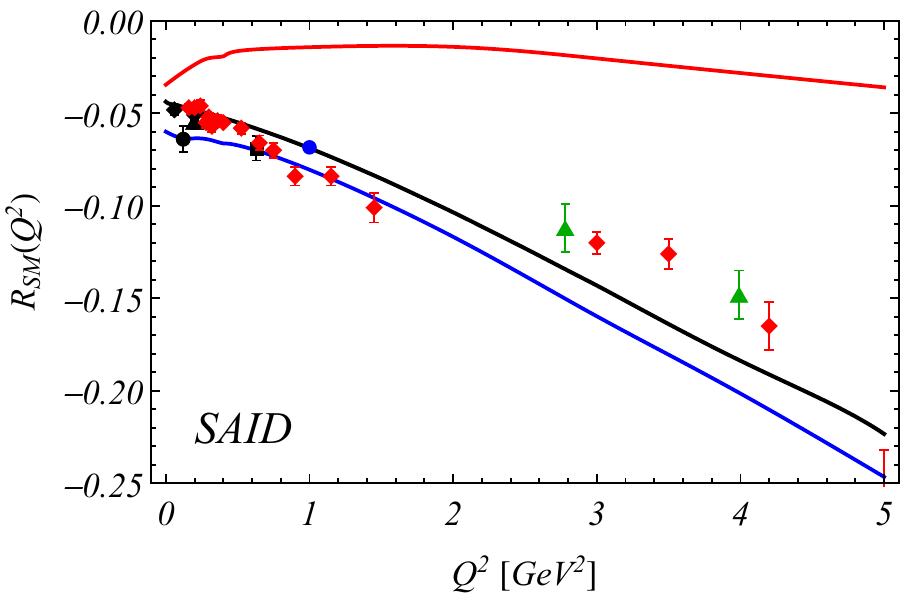}
\end{center}
\vspace{3mm} \caption{\label{fig:3} Magnetic transition form factor in the notation of Ash divided by the dipole form
factor $G_M/G_D$ and $E/M$ and $S/M$ ratios compared at the BW and pole position. The black curves
show the real BW results and the blue and red lines show real and imaginary parts of the complex
pole form factors. The left column shows the results with the Mainz-MAID analysis and the right
column with the GWU-SAID analysis.
%
%
The data points for $G_M$ are from Refs: \cite{Beck00} (black circle), \cite{Bartel} (black
squares), \cite{Stein} (blue circles), \cite{Fro99} (green triangles) and \cite{Azn09} (red
diamonds);
for $E/M$: \cite{Beck00} (black circle), \cite{Stave:2008tv} (black diamonds), \cite{Mer01} (black
triangle), \cite{Got00,Ban03} (black square), \cite{Kelly05} (blue circle), \cite{Fro99} (green
triangles) and \cite{Azn09} (red diamonds);
and for $S/M$: \cite{Stave:2008tv} (black diamonds), \cite{Pos01} (black circle), \cite{Elsner06}
(black triangle), \cite{Got00,Ban03} (black square), \cite{Kelly05} (blue circle), \cite{Fro99}
(green triangles) and \cite{Azn09} (red diamonds).
}

\end{figure*}

In Fig. \ref{fig:3}, we compare the quantities $G_M/G_D$, and the $E/M$ and $S/M$ ratios as
functions of $Q^2$, where $G_D = (1+Q^2/b^2 )^{-2}$, with $b^2$ = 0.71 (GeV/c)$^2$. Here also, the
MAID and SAID results for $G_M/G_D$ agree very closely, with only a small difference between the BW
values and real parts of the pole quantities. This pole behavior has also been displayed, over a
smaller $Q^2$ range, in the analysis of Ref.~\cite{suzu10}. The MAID and SAID BW results also agree
well with the available single-$Q^2$ analyses of the $E/M$ ratio. These plots give no indication of
a cross-over to positive $E/M$ values, as expected from Ref.~\cite{carlson}. Previously, both the
MAID (2003)~\cite{maid2003} and SAID (2002)~\cite{vpi2002} fits had found indications for a cross
over. This trend has disappeared with the incorporation of new and more precise measurements. The
$S/M$ ratios of the MAID and SAID analyses display the only qualitative difference in $Q^2$
variation. Here also the BW and real part of the pole behavior is similar, with the SAID (pole and
BW) curves tending to approximately follow the behavior of the single-$Q^2$ fits, whereas the MAID
trend is for a slower $Q^2$ variation. We note that in the 2003 MAID analysis~\cite{maid2003}, the
$S/M$ ratio was found to have a more rapid $Q^2$ variation, following the trend of existing
single-$Q^2$ values.

\begin{figure*}[h]
\begin{center}
\includegraphics[width=7.5cm]{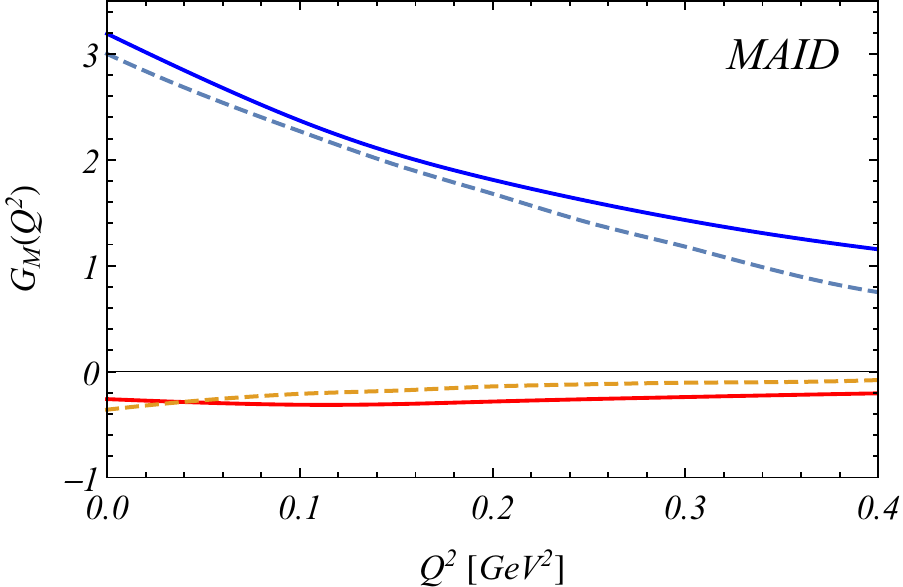}
\hspace*{0.5cm}
\includegraphics[width=7.5cm]{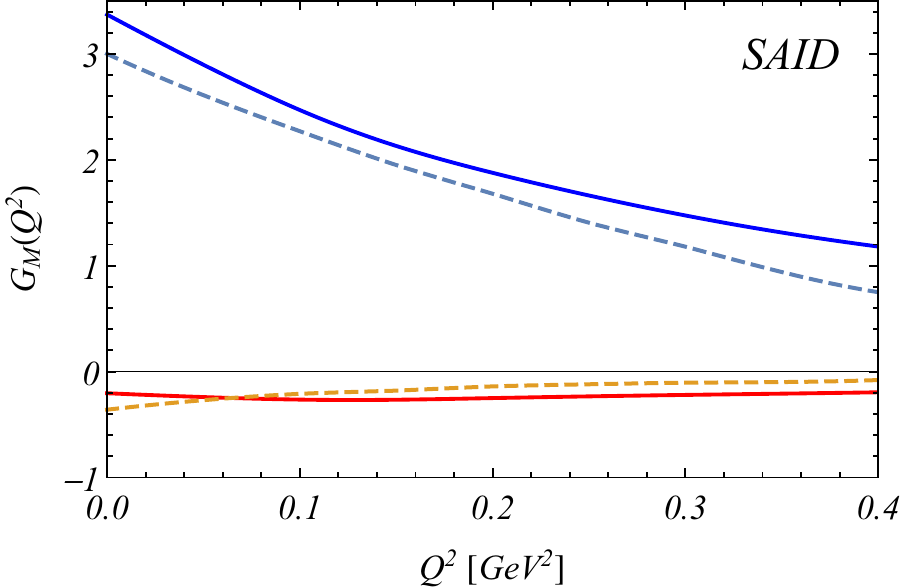}

\vspace*{0.5cm}
\includegraphics[width=7.5cm]{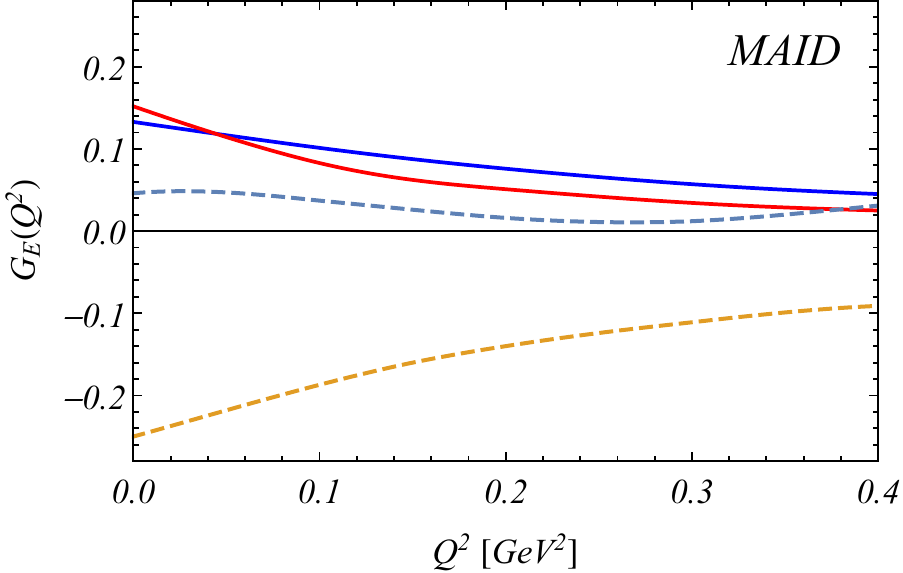}
\hspace*{0.5cm}
\includegraphics[width=7.5cm]{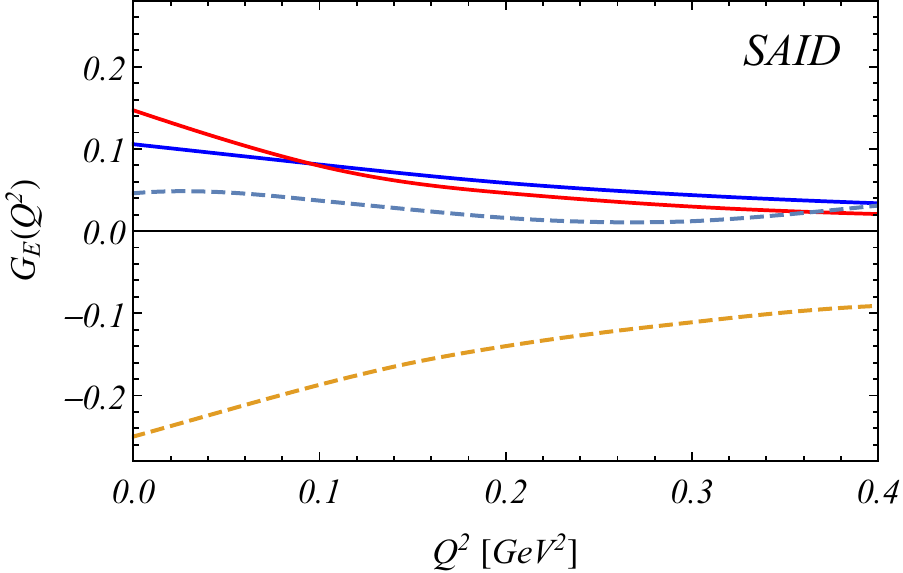}

\vspace*{0.5cm}
\includegraphics[width=7.5cm]{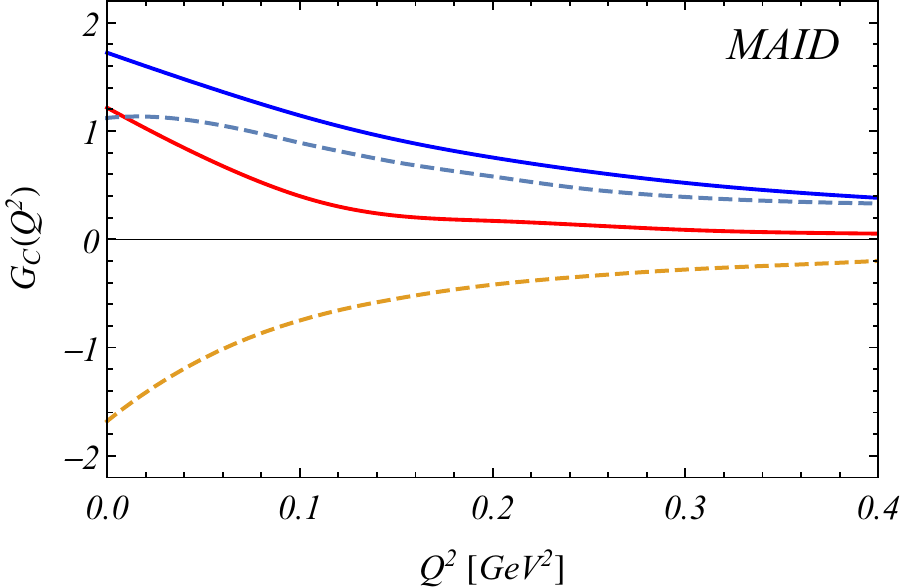}
\hspace*{0.5cm}
\includegraphics[width=7.5cm]{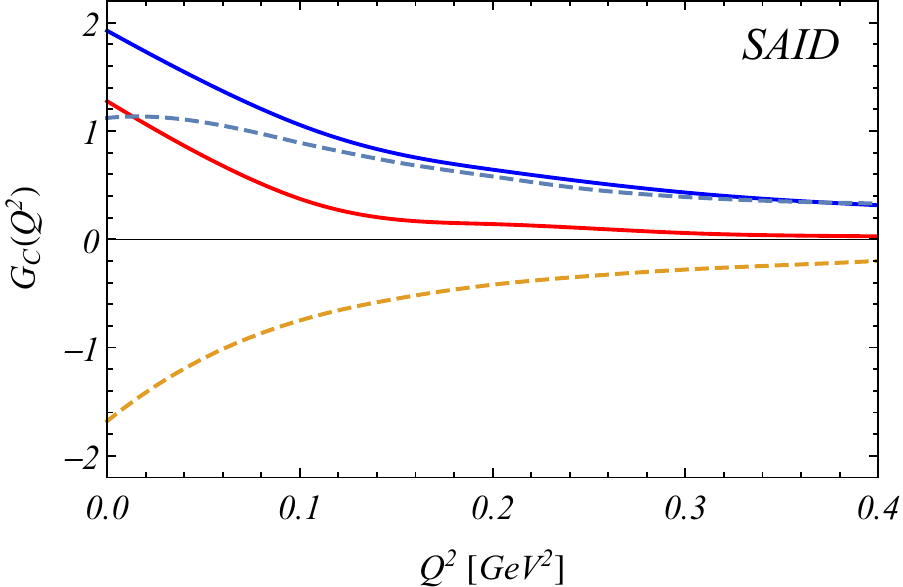}
\end{center}
\vspace{3mm} \caption{\label{fig:4} Magnetic, electric and charge transition form factors compared
with the Heavy Baryon chiral effective field theory of Gail and Hemmert~\cite{Gail} at low $Q^2$.
The blue and red lines show real and imaginary parts of the complex pole form factors obtained from
MAID and SAID. The dashed lines are the HBChEFT calculations.}
\end{figure*}

For low values of $Q^2$, we can also compare to the expectations from chiral effective
theory~\cite{Gail}. In Fig.~\ref{fig:4}, the MAID and SAID quantities from Fig.~\ref{fig:3} are
compared to the predictions of Gail and Hemmert, Ref.~\cite{Gail} over a restricted $Q^2$ range.
The range of applicability of their approach was estimated to about $Q^2_{max}\approx 0.2$~GeV$^2$.
Due to the lack of data at the pole position, single-Q$^2$ data extracted at the BW position were
used to determine the parameters of their approach. The result is a qualitatively good agreement
between the real parts of pole-valued quantities, especially for the dominant magnetic transition,
where even the imaginary part is reasonably described. However, this is not the case for $G_E$ and
$G_C$. The real parts of these transitions are still in a moderate agreement, but the imaginary
parts are off even by a different sign. This is not too surprising because the imaginary parts
strongly depend on the parameters used for the pion loop integrals. A revised relativistic ChPT
calculation in the complex mass scheme~\cite{scherer2016} is in progress and may shed light on this
issue.

\begin{table*}[ht]
\begin{tabular}{|l|rrr|rrr|}\hline
& \multicolumn{3}{|c|}{MAID Values} & \multicolumn{3}{|c|}{SAID Values}\\
& BW & \multicolumn{2}{c|}{pole} & BW & \multicolumn{2}{c|}{pole}\\
\hline
$G_M$ & $2.97$ & $3.20$ & $-4.7^{\circ} $ & $3.11$ & $3.38$ & $-3.5^{\circ} $ \\
$G_E$ & $0.064$ & $0.202$ & $49^{\circ} $ & $0.051$ &$0.181$ & $54^{\circ} $ \\
$G_C$ & $1.18$ & $2.11$ & $35^{\circ} $   & $1.30$ & $2.31$ & $34^{\circ} $ \\
\hline
$R_{EM}$ & $-0.022$ & $-0.063$ & $53^{\circ} $ & $-0.016$ & $-0.054$ & $58^{\circ} $ \\
$R_{SM}$ & $-0.042$ & $-0.067$ & $33^{\circ} $ & $-0.044$ & $-0.069$ & $30^{\circ} $ \\
\hline
$A_{1/2}$ & $-0.131$ & $-0.131$ & $-20^{\circ} $ &  $-0.139$ & $-0.142$ & $-18^{\circ} $ \\
$A_{3/2}$ & $-0.247$ & $-0.261$ & $-7.7^{\circ} $ & $-0.258$ & $-0.273$ & $-6.8^{\circ} $ \\
$S_{1/2}$ & $ 0.016$ & $ 0.027$ & $22^{\circ} $ &   $ 0.018$ & $ 0.030$ & $ 21^{\circ} $ \\
\hline
\end{tabular}
\caption{\label{tab:pole1} Magnetic, electric and charge transition form factors, $E/M$, $S/M$
ratios and photon decay amplitudes at $Q^2=0$ for the Breit-Wigner and for the pole position
compared between MAID and SAID solutions. The BW parameters used for the conversion factor are:
$M_\Delta=1232$~ MeV and $\Gamma_\pi = \Gamma_r = 115$~MeV, and the pole parameters are: $W_p=(1210
- 50 i)$~MeV and $Res_{\pi N}=50\,e^{-i 47^\circ}$. The form factors and ratios are dimensionless
and the photon decay amplitudes are given in units of $10^{-3}/GeV^{1/2}$. For the complex values
at the pole position, we give absolute values with the same sign as for the BW values and a phase.}
\end{table*}

%

\begin{acknowledgments}
This work was supported in part by the U.S. Department of Energy
Grant DE-SC0014133, the Deutsche Forschungsgemeinschaft (SFB 1044),
and the RFBR grant 13-02-00425. M.D. would like to the the National Science Foundation (CAREER grant No.
PHY-1452055 and PIF grant No. 1415459) for support.

\end{acknowledgments}

\clearpage



\begin{thebibliography}{AA}

\bibitem{Workman:2013rca}
  R.~L.~Workman, L.~Tiator and A.~Sarantsev,
  Phys.\ Rev.\ C {\bf 87}, no. 6, 068201 (2013).

\bibitem{pdg}
  K.~A.~Olive {\it et al.} [Particle Data Group Collaboration],
  Chin.\ Phys.\ C {\bf 38}, 090001 (2014);
  J.~Beringer {\it et al.},
  Phys.\ Rev.\ D\ \textbf{86}, 010001 (2012).

\bibitem{bnga}
  A.V. Anisovich, R. Beck, E. Klempt, V.A. Nikonov, A.V. Sarantsev, U. Thoma,
  Eur. Phys. J. \textbf{A48}, 15 (2012).

\bibitem{Doring:2009yv}
  M.~D\"oring, C.~Hanhart, F.~Huang, S.~Krewald and U.-G.~Mei{\ss}ner,
  Nucl.\ Phys.\ A {\bf 829}, 170 (2009).

\bibitem{ceci11}
S. Ceci, M. D\"oring, C. Hanhart, S. Krewald, U.-G.Meissner, A. Svarc,
Phys. Rev. C \textbf{84}, 015205 (2011);

\bibitem{masj11}
P. Masjuan, AIP Conf. Proc. 1343, 334 (2011);

\bibitem{yang11}
S.N. Yang, Chin. J. Phys. 49, 1157 (2011);

\bibitem{tiat10}
L. Tiator, S.S. Kamalov, S. Ceci, G.Y. Chen, D. Drechsel, A. Svarc,
S.N. Yang, Phys. Rev. C \textbf{82}, 055203 (2010);

\bibitem{suzu10}
N. Suzuki, T. Sato, T.-S. H. Lee, Phys. Rev. C \textbf{82}, 045206
(2010);

\bibitem{svar12}
A. Svarc, M. Hadzimehmedovic, H. Osmanovic, J. Stahov, arXiv:1212.1295[nucl-th].

\bibitem{Kamano:2013iva}
  H.~Kamano, S.~X.~Nakamura, T.-S.~H.~Lee and T.~Sato,
  Phys.\ Rev.\ C {\bf 88}, no. 3, 035209 (2013).


\bibitem{Ronchen:2015vfa}
  D.~R\"onchen, M.~D\"oring, H.~Haberzettl, J.~Haidenbauer, U.-G.~Mei{\ss}ner and K.~Nakayama,
  Eur.\ Phys.\ J.\ A {\bf 51}, 70 (2015).

\bibitem{Ronchen:2014cna}
  D.~R\"onchen {\it et al.},
  Eur.\ Phys.\ J.\ A {\bf 50}, 101 (2014)
  Erratum: [Eur.\ Phys.\ J.\ A {\bf 51}, 63 (2015)].


\bibitem{Gail}
T.A.~Gail and T.R.~Hemmert, Eur. Phys. J. A \textbf{28}, 91 (2006).

\bibitem{Pascalutsa:2005vq}
  V.~Pascalutsa and M.~Vanderhaeghen,
  Phys.\ Rev.\ D {\bf 73}, 034003 (2006).

\bibitem{Bernard:2005fy}
  V.~Bernard, T.~R.~Hemmert and U.~G.~Mei{\ss}ner,
  Phys.\ Lett.\ B {\bf 622}, 141 (2005).

\bibitem{Jido:2007sm}
  D.~Jido, M.~D\"oring and E.~Oset,
  Phys.\ Rev.\ C {\bf 77}, 065207 (2008).

\bibitem{Doring:2007rz}
  M.~D\"oring,
  Nucl.\ Phys.\ A {\bf 786}, 164 (2007).

\bibitem{Doring:2010rd}
 M.~D\"oring, D.~Jido and E.~Oset,
  Eur.\ Phys.\ J.\ A {\bf 45}, 319 (2010)

\bibitem{Aznauryan:2016wwm}
  I.~G.~Aznauryan and V.~D.~Burkert,
  arXiv:1603.06692 [hep-ph].

\bibitem{Ramalho:2015qna}
  G.~Ramalho, M.~T.~Pe\~na, J.~Weil, H.~van Hees and U.~Mosel,
  Phys.\ Rev.\ D {\bf 93}, 033004 (2016).

\bibitem{aznauryan1}
I. Aznauryan and V.D. Burkert,
Phys. Rev. C \textbf{92}, 035211 (2015).

\bibitem{segovia}
J. Segovia, I.C. Clo\"et, C.D. Roberts, S.M. Schmidt,
Few Body Syst. \textbf{55}, 1185 (2014).

\bibitem{Segovia:2013uga}
  J.~Segovia, C.~Chen, I.C.~Clo\"et, C.D.~Roberts, S.~M.~Schmidt and S.~Wan,
  Few Body Syst.\  {\bf 55}, 1 (2014).

\bibitem{Ronniger:2012xp}
  M.~Ronniger and B.~C.~Metsch,
  Eur.\ Phys.\ J.\ A {\bf 49}, 8 (2013).


\bibitem{Sanchis-Alepuz:2013iia}
  H.~Sanchis-Alepuz, R.~Williams and R.~Alkofer,
  Phys.\ Rev.\ D {\bf 87}, no. 9, 096015 (2013).

\bibitem{Ramalho:2012ng}
  G.~Ramalho and M.~T.~Pe\~na,
  Phys.\ Rev.\ D {\bf 85}, 113014 (2012).

\bibitem{Santopinto:2012nq}
  E.~Santopinto and M.~M.~Giannini,
  Phys.\ Rev.\ C {\bf 86}, 065202 (2012).


\bibitem{Aznauryan:2011qj}
  I.~G.~Aznauryan and V.~D.~Burkert,
  Prog.\ Part.\ Nucl.\ Phys.\  {\bf 67}, 1 (2012).


\bibitem{Santopinto:2010zz}
  E.~Santopinto, A.~Vassallo, M.~M.~Giannini and M.~De Sanctis,
  Phys.\ Rev.\ C {\bf 82}, 065204 (2010).



\bibitem{ebac}
B. Juli\'a-D\'iaz {\it et al.}, Phys. Rev. C \textbf{80}, 025207 (2009).

\bibitem{ungaro}
M. Ungaro {\it et al.}, Phys. Rev. Lett. \textbf{97}, 112003 (2006).

\bibitem{Fiolhais:1996bp}
  M.~Fiolhais, B.~Golli and S.~Sirca,
  Phys.\ Lett.\ B {\bf 373}, 229 (1996).

\bibitem{Pascalutsa:2006up}
  V.~Pascalutsa, M.~Vanderhaeghen and S.~N.~Yang,
  Phys.\ Rept.\  {\bf 437}, 125 (2007).

\bibitem{Alexandrou:2013ata}
 C.~Alexandrou, J.~W.~Negele, M.~Petschlies, A.~Strelchenko and A.~Tsapalis,
  Phys.\ Rev.\ D {\bf 88}, no. 3, 031501 (2013)

\bibitem{Alexandrou:2010uk}
 C.~Alexandrou, G.~Koutsou, J.~W.~Negele, Y.~Proestos and A.~Tsapalis,
  Phys.\ Rev.\ D {\bf 83}, 014501 (2011)

\bibitem{Agadjanov:2014kha}
A.~Agadjanov, V.~Bernard, U.-G.~Mei{\ss}ner and A.~Rusetsky,
  Nucl.\ Phys.\ B {\bf 886}, 1199 (2014)

\bibitem{L+P2013} A. \v{Svarc}, M. Had\v{z}imehmedovi\'{c}, H. Osmanovi\'{c}, J. Stahov, L. Tiator, and R. L. Workman, Phys. Rev. \textbf{C88}, 035206 (2013).
\bibitem{L+P2014} A. \v{Svarc}, M. Had\v{z}imehmedovi\'{c}, R. Omerovi\'{c}, H. Osmanovi\'{c}, and J. Stahov,  Phys. Rev. \textbf{C89}, 45205 (2014).
\bibitem{L+P2015} A. \v{Svarc}, M. Had\v{z}imehmedovi\'{c}, H. Osmanovi\'{c}, J. Stahov,  and R. L. Workman, Phys. Rev. \textbf{C91}, {015207} (2015).
\bibitem{L+P2014a} A. \v{Svarc}, M. Had\v{z}imehmedovi\'{c}, H. Osmanovi\'{c}, J. Stahov, L. Tiator, and R. L. Workman, Phys. Rev. \textbf{C89}, 065208 (2014).
\bibitem{L+P2016} A. \v{Svarc}, M. Had\v{z}imehmedovi\'{c}, H. Osmanovi\'{c}, J. Stahov, L. Tiator, and R. L. Workman, Phys. Lett. \textbf{B755}, 452-455 (2016).

\bibitem{bm}
C. Becchi and G. Morpurgo, Phys. Lett. \textbf{17}, 352 (1965).

\bibitem{carlson}
C.E.~Carlson, Phys.\ Rev.\ D\ \textbf{34}, 2704 (1986).

\bibitem{Ash67} W.~W. Ash et al., Phys. Lett. B {\bf 24} (1967) 165.

\bibitem{Jon73} H.~F. Jones and M.~D. Scadron, Annals Phys. {\bf 81}, 1 (1973).

\bibitem{Tiator:2011pw}
  L.~Tiator, D.~Drechsel, S.~S.~Kamalov and M.~Vanderhaeghen,
  Eur.\ Phys.\ J.\ ST {\bf 198}, 141 (2011).

\bibitem{Devenish:1975jd}
  R.~C.~E.~Devenish, T.~S.~Eisenschitz, J.~G.~K\"orner,
  Phys.\ Rev.\  {\bf D14 } (1976)  3063.

\bibitem{maid2007}
D. Drechsel, S.S. Kamalov, L. Tiator,
Nucl. Phys. \textbf{A34}, 69 (2007).

\bibitem{Doring:2009bi}
  M.~D\"oring, C.~Hanhart, F.~Huang, S.~Krewald and U.-G.~Mei{\ss}ner,
  Phys.\ Lett.\ B {\bf 681}, 26 (2009).

\bibitem{Hoehler93} G. H\"{o}hler, $\pi$N Newsletter {\textbf 9}, 1 (1993).

\bibitem{Cutkosky79} R. E. Cutkosky, C. P. Forsyth, R. E. Hendrick, and R.L. Kelly, Phys. Rev. {\bf D 20}, 2839 (1979).

\bibitem{Ceci2008} S. Ceci, J. Stahov, A. \v{S}varc, S. Watson, and B. Zauner, Phys. Rev. {\textbf D 77}, 116007 (2008).

\bibitem{web}
Amplitudes are available from the MAID site: \hbox{http://portal.kph.uni-mainz.de/MAID//} and
from the SAID site: \hbox{http://gwdac.phys.gwu.edu}.

\bibitem{maid2003}
L. Tiator, D. Drechsel, S.S. Kamalov, S.N. Yang,
Eur. Phys. J. A \textbf{17}, 357 (2003).

\bibitem{vpi2002}
R.A. Arndt, W. Briscoe, I. Strakovsky, R. Workman,
p.234, proceedings of the Workshop on the Physics of Excited Nucleons (NSTAR2002),
Pittsburgh, PA, 2002. Ed. S.A. Dytman and E.S. Swanson (World Scientific, 2003).

\bibitem{scherer2016}
M. Hilt, T. Bauer, S. Scherer, L. Tiator, in preparation.

\bibitem{Stave:2008tv}
  S.~Stave et al. [ A1 Collaboration ],
  Phys.\ Rev.\  {\bf C78 } (2008)  025209.

\bibitem{Azn09}
  I.~G.~Aznauryan et al. [ CLAS Collaboration ],
  Phys.\ Rev.\  {\bf C80 } (2009)  055203.

\bibitem{Bartel} W. Bartel {et al.}, Phys. Lett. {\bf 28B} (1968) 148

\bibitem{Stein} S. Stein et al., Phys. Rev. {\bf D12} (1975) 1884.

\bibitem{Pos01} Th.~Pospischil {et al.}, Phys. Rev. Lett. {\bf 86} (2001) 2959.

\bibitem{Elsner06} D.~Elsner {et al.}, Eur. Phys. J. A {\bf 27} (2006) 91.

\bibitem{Stave06} S.~Stave {et al.}, Eur. Phys. J. A {\bf 30} (2006) 471.

\bibitem{Got00} R.~W.~Gothe, Prog. Part. Nucl. Phys. {\bf 44} (2000) 185.
\bibitem{Ban03} T.~Bantes, PhD thesis, Bonn 2003.


\bibitem{Mer01} C.~Mertz {et al.}, Phys. Rev. Lett. {\bf 86} (2001) 2963.

\bibitem{Beck00} R.~Beck et al., Phys.\ Rev.\  C {\bf 61} (2000) 035204.

\bibitem{Kelly05}J.~J.~Kelly {et al.}, Phys. Rev. Lett. {\bf 95} (2005) 102001
and Phys. Rev. C {\bf 75} (2007), 025201.

\bibitem{Fro99} V.~V.~Frolov {et al.}, Phys. Rev. Lett. {\bf 82} (1999) 45.


\end{thebibliography}
\end{document}